\documentclass[prc,aps,twocolumn,floatfix,superscriptaddress]{revtex4}
\usepackage{bm}
\usepackage{amsmath}
\usepackage{amsfonts}
\usepackage{amssymb}
\usepackage{graphicx}

\begin{document}

\title{Extrapolation of scattering data to the negative-energy region. Application to the $p-^{16}$O system}

\author{L. D. Blokhintsev}
\email{blokh@srd.sinp.msu.ru}
\affiliation{Skobeltsyn Institute of Nuclear Physics, Lomonosov Moscow State University, Moscow 119991, Russia}
\affiliation{Pacific National University, Khabarovsk 680035, Russia}
\author{A. S. Kadyrov}
\email{a.kadyrov@curtin.edu.au}
\affiliation{Curtin Institute for Computation and Department of Physics and Astronomy, Curtin University, GPO Box U1987, Perth, WA 6845, Australia}
\author{A. M. Mukhamedzhanov}
\email{akram@comp.tamu.edu}
\affiliation{Cyclotron Institute, Texas A\&M University, College Station, Texas 77843, USA}
\author{D. A. Savin} 
\email{rhdp@mail.ru}
\affiliation{Skobeltsyn Institute of Nuclear Physics, Lomonosov Moscow State University, Moscow 119991, Russia}

\begin{abstract}
The problem of analytic continuation of the scattering data to the negative-energy region  to obtain information on asymptotic normalization coefficients (ANCs) of bound states is discussed.  It is shown that  a recently   suggested $\Delta$ method [O.L.Ram\'{\i}rez Su\'arez and J.-M. Sparenberg, Phys. Rev. C {\bf 96}, 034601 (2017)]  is not strictly correct in the mathematical sense since it is not an analytic continuation of a partial-wave scattering amplitude  to the region of negative energies.  However, it can be used for practical purposes for sufficiently large charges and masses of colliding particles. Both the $\Delta$ method and the standard method of continuing of the effective range function are applied to the $p-^{16}$O system which is of interest for nuclear astrophysics. The ANCs  for the ground
$5/2^+$ and excited $1/2^+$ states of $^{17}$F are determined.

\end{abstract}

\maketitle

\section{Introduction}

Using scattering data may give valuable information on the features of bound states, particularly on asymptotic normalization coefficients (ANCs), which, in contrast to binding energies, cannot be directly measured. The ANCs are fundamental nuclear characteristics that are important, for example, for evaluating cross sections of peripheral astrophysical nuclear reactions \cite{MukhTim,Xu,MukhTr,reviewpaper}.  One of the direct ways  to extract ANCs from experimental data is the analytic continuation in the energy plane of the partial-wave elastic scattering amplitudes, obtained by the phase-shift analysis,  to the pole corresponding to a bound state. Such a procedure, in contrast to the method of constructing optical potentials fitted to scattering data, allows one to circumvent an ambiguity problem associated with the existence of phase-equivalent potentials \cite{BlEr,BlOrSa}. 

The conventional procedure for such extrapolation is the analytic approximation of the experimental values of the  effective-range function 
(ERF) $K_l(E)$ with the subsequent continuation to the pole position ($l$ is the orbital angular momentum). The ERF method has been successfully employed to determine the ANCs for bound (as well as resonant) nuclear states in a number of works (see, e.g. \cite{BKSSK,SpCaBa,IrOr} and references therein). 

The ERF is expressed in terms of scattering phase shifts. In the case of charged particles, 
the ERF for the short-range interaction should be modified. Such modification 
generates additional terms in the ERF. These terms depend only on the Coulomb interaction and may far exceed, in the absolute value, the informative part of the ERF containing the phase shifts. This fact may hamper the practical procedure of the analytic continuation and affect its accuracy. It was suggested in Ref.~\cite{Sparen} to use for the analytic continuation the quantity $\Delta_l(E)$ (which is defined below in Section II) rather than the ERF $K_l(E)$. The $\Delta_l(E)$ function does not contain the pure Coulomb terms. We  call the continuation method, which uses  the $\Delta$ function, the $\Delta$ method.  In \cite{Sparen2} this method is called the reduced ERF method.

Note that the validity of employing $\Delta_l(E)$ was not obvious, which resulted in some discussion. The authors of Refs. 
\cite{OrIrNa,IrOr1} claimed that they proved the mathematical correctness of the $\Delta$ method. However, this assertion contradicts the results of Refs.\cite{BKMS2,Sparen2}. 

In the present work, we consider the question of the validity and applicability of the $\Delta$ method. It is shown that  the 
$\Delta$ method in the strict mathematical sense is not an analytic continuation of a partial-wave scattering amplitude  to the region of negative energies, however, it can be used for practical purposes for sufficiently large charges and masses of colliding particles. Then both ERF and $\Delta$ methods are employed to analyze the $p-^{16}$O system and determine the ANCs for ground and excited states of $^{17}$F in the $p-^{16}$O channel. Note that the knowledge of these ANCs is important for evaluating  the astrophysical $S$-factor of the $^{16}\mathrm{O}(p,\gamma)^{17}$F reaction which is one of the processes of the CNO cycle of nucleosynthesis in stars \cite{Gagliardi}. The analysis is based on using the experimental phase shifts with corresponding experimental errors.  It is demonstrated here that the extrapolation of the elastic scattering data to the bound state poles provides 
a practical method to determine the ANCs. The ANCs, which are determined  by the extrapolation of the elastic scattering data to the bound state poles, can be called experimental
ANCs because they are obtained from  experimental data. 

The paper is organized as follows. Section II contains the general formalism of the elastic scattering for the superposition of a short-range and the Coulomb interactions which is necessary for the subsequent discussion. The validity and applicability of the $\Delta$ method is discussed in Sect. III.  Experimental $p-^{16}$O phase shifts are used to determine the ANCs for $^{17}$F in Sect. IV.  Throughout the paper we use the system of units in which $\hbar=c=1$ .

\section{Basic formalism}

In this section, we recapitulate basic equations which are necessary for the subsequent discussion. 
The Coulomb-nuclear  amplitude of the elastic scattering of particles 1 and 2 is given by
\begin{equation}\label{fNC}
f_{NC}({\rm {\bf  k}})=\sum_{l=0}^\infty(2l+1)\exp(2i\sigma_l)\frac{\exp(2i\delta_l)-1}{2ik}P_l(\cos\theta).
\end{equation}
Here ${\rm {\bf k}}$ is the relative momentum of particles 1 and 2, $\theta$ is the center of mass  (c.m.)  scattering angle,   
$\sigma_l=\arg\,\Gamma(l+1+i\eta)$  
 and $\delta_l$ are the pure Coulomb and Coulomb-nuclear phase shifts, respectively, and $\Gamma(z)$ is the Gamma function. The Coulomb  parameter for the 1+2 scattering state is given as
\begin{equation}\label{eta}
\eta =Z_1Z_2e^2\mu/k,
\end{equation}
where the relative momentum $k$ is related to the relative energy of these particles $\,E$ by  $\,k=\sqrt{2\mu E}$,
$\mu=m_1m_2/(m_1+m_2)$, $m_i$ and $Z_ie$  are the mass and the electric charge of particle $i$, $i=1,2$. 
 
The behavior of the Coulomb-nuclear partial-wave amplitude $f_l=(\exp(2i\delta_l)-1)/2ik$ is irregular near 
$E=0$. Therefore, one can introduce renormalized Coulomb-nuclear partial-wave amplitude $\tilde f_l$ \cite{Hamilton,BMS,Konig} according to
\begin{equation}\label{renorm}
\tilde f_l=\exp(2i\sigma_l)\,\frac{\exp(2i\delta_l)-1}{2ik}\,\left[\frac{l!}{\Gamma(l+1+i\eta)}\right]^2e^{\pi\eta}.
\end{equation}
Eq.~(\ref{renorm}) can be rewritten as 
\begin{equation}\label{renorm1}
\tilde f_l=\frac{\exp(2i\delta_l)-1}{2ik}C_l^{-2}(\eta),
\end{equation}
where $C_l(\eta)$ is the Coulomb penetrability factor (or Gamow factor) determined by
\begin{align}\label{C}
C_l(\eta)&=\left[\frac{2\pi\eta}{\exp(2\pi\eta)-1}v_l(\eta)\right]^{1/2}, \\ %\quad 
v_l(\eta)&=\prod_{n=1}^{l}(1+\eta^2/n^2)\;(l>0),\quad v_0(\eta)=1.
\end{align}
It was shown in Ref. \cite{Hamilton} that  the
analytic properties of ${\tilde f}_{l}$ on the physical sheet of $E$  are analogous to the ones of the partial-wave scattering amplitude for the short-range potential and it can be analytically continued into the negative energy region.

The amplitude $\tilde f_l$ can be expressed in terms of the Coulomb-modified ERF $K_l(E)$ \cite{Hamilton, Konig}  by
\begin{align} 
\label{fK}
\tilde f_l&=\frac{k^{2l}}{K_l(E)-2\eta k^{2l+1}h(\eta)v_l(\eta)}\\ %\nonumber 
&=\frac{k^{2l}}{k^{2l+1}C_l^2(\eta)(\cot\delta_l-i)} \\ %\nonumber 
&=\frac{k^{2l}}{v_l^2 k^{2l}\Delta_l(E)-ik^{2l+1}C_l^2(\eta)},
\label{fK3}
\end{align}   
where
\begin{align}\label{scatfun}
K_l(E)&= k^{2l+1} \left[ C_l^2(\eta)(\cot\delta_l-i) + 2 \eta h(k)v_l(\eta) \right],\\ %\nonumber
h(\eta) &= \psi(i\eta) + \frac{1}{2i\eta}-\ln(i\eta), \\  
\Delta_l(E)&=kC_0^2(\eta)\cot\delta_l, 
\label{Deltal}
\end{align}
and $\psi(x)$ is the digamma function. $\Delta_l(E)$ is the $\Delta$ function introduced in \cite{Sparen}.  It was shown in \cite{Hamilton} that function $K_l(E)$ defined by Eq. (\ref{scatfun}) is analytic near $E=0$ and can be expanded into a Taylor series in $E$. In the absence of the Coulomb interaction ($\eta=0$) $K_l(E)=k^{2l+1}\cot\delta_l(k)$.

If the $1+2$ system has  a  bound state $3=(1\,2)$ with the binding energy $\varepsilon>0$ in the partial wave $l$, then the amplitude $\tilde f_l$ has a pole at $E=-\varepsilon$.           The residue of $\tilde f_l$ at this point is expressed in terms of the ANC
$C^{(l)}_{3\to 1+2}$ \cite{BMS} as
\begin{align}\label{res2}
{\rm res}\tilde f_l(E)|_{E=-\varepsilon}&=\lim_{\substack{E\to -\varepsilon}}[(E+\varepsilon)\tilde f_l(E)] \\
&=
-\frac{1}{2\mu}\left[\frac{l!}{\Gamma(l+1+\eta_b)}\right]^2 \left[C^{(l)}_{3\to 1+2}\right]^2,
\label{res22}
\end{align}
where $\eta_b=Z_1Z_2e^2\mu/\kappa$ is the Coulomb  parameter for the bound state 3  and $\kappa$ is the bound-state wave number.

\section{On the validity and applicability of the $\Delta$ method}

  In this Section, we discuss general properties of the $\Delta$ method suggested in Ref.~\cite{Sparen}. Within this method, one uses 
	for the analytic continuation the quantity $\Delta_l(E)$ given by Eq. (\ref{Deltal}) rather than the ERF $K_l(E)$ of Eq. (\ref{scatfun}). The reasons for introducing the $\Delta$ method  are outlined above in the introduction. However, the validity of employing $\Delta_l(E)$ was not obvious since $\Delta_l(E)$, in conrast to $K_l(E)$, possesses an essential singularity at $E=0$.

For brevity, the subsequent formulas  in this section are written for the $s$-wave case and index $l=0$ is omitted. Nevertheless, all reasonings are valid for arbitrary $l$.

Consider the partial-wave amplitude  $\tilde f$. We write
\begin{equation}
\tilde f(E)=\frac{1}{D(E)},%\quad D(E)=kC^2(\eta)\cot\delta-ikC^2(\eta)\equiv\Delta(E)-ig(E)
\label{tilde}
\end{equation}
where
\begin{equation}
D(E)=kC^2(\eta)\cot\delta-ikC^2(\eta)\equiv\Delta(E)-ig(E).
\label{D}
\end{equation}
If the Coulomb interaction is switched off, then
\begin{equation}\label{switch}
C^2(\eta)=1,\quad D(E)=k\cot\delta-ik.
\end{equation}
Denote $E=E_+$ if $E>0$ and $E=E_-$ if $E<0$. 

Note that $ig(E_+)$ is pure imaginary. At $E=0$ the latter has the essential and square-root singularities. On the other hand, $ig(E_-)$ is complex. Also,
$\mathrm {Im}\, \Delta(E_+)=0$ and at $E=0$ $\Delta(E)$ possesses the essential singularity. For $\,E_- \,$ the imaginary parts of  $\,\Delta(E_-)\,$ and  $\,ig(E_-)\,$ cancel each other and  the essential singularity in Eq. (\ref{tilde}) is cancelled as well.  As a result,  $\mathrm  {Im}\, D(E_-)=0$.

It should be emphasized that $D(E_+)$ and $D(E_-)$ are different parts of the same analytic function.  The analytic continuation of 
$D(E)$ from  $E_+$ to  $E_-$ implies, as in the case of neutral particles (see Eq. (\ref{switch})), that the whole function  $\,D(E_+)=\Delta(E_+)-ig(E_+)\,$ should be continued rather than only $\,\Delta(E_+)$.  Note that in the $\,\Delta\,$ method $\,\Delta(E_+)\,$  is approximated by polynomials or rational functions in $E$ and then continued to $E_-$ where the approximated $\Delta(E_-)$ is equated to the whole denominator  $D(E_-)$ and the position of the pole of $\tilde f(E)$ corresponding to a bound state is determined by the condition $\,\Delta(E_-)=0$.  
	
Obviously, such a procedure cannot be regarded as mathematically correct.  In particular, it does not reproduce the square-root singularity (the normal threshold) of $\tilde f(E)$ at $E=0$. The analytic continuation of $\Delta(E_-)$ thus obtained back to $E_+$ results in a wrong equation $\,\mathrm{Im}\,\tilde f(E_+)=0.\,$  

We note, however, that  in the case of a purely short-range interaction  Im $D(E_+)$  decreases as $\sqrt E$ at $E\to 0$ and in the presence of a repulsive Coulomb potential it decreases exponentially:
\begin{equation}
\mathrm{Im}\,\, D(E_+)|_{E\to 0}\sim e^{-\gamma/\sqrt E}, \quad \gamma= \pi\sqrt{2\mu}Z_1Z_2e^2.
\end{equation}
And not only $\mathrm{Im}\,D(E_+)$ but all its derivatives  tend to zero at $E\to +0$ as distinct from the case of neutral particles scattering.  
Hence, in the presence of the Coulomb ineraction there is a range of values of $E$  in the vicinity of $E=0$ in which one can neglect  Im $D(E_+)$ and consider that $D(E_+)\approx \Delta(E_+)$. Within this range $D(E_+)$ can be approxinated by a polynomial or a rational function of $E$ and then continued to $E_-$. The size of this range can be qualitatively determined by the condition
\begin{equation}
|E|\ll \gamma^2.
\end{equation} 
The problem of the validity and applicability of the $\Delta$ method was discussed in Refs. \cite{BKMS1,BKMS2,Sparen2}. It was stated in Refs. \cite{BKMS2,Sparen2} that the $\Delta$ method can be employed to obtain information on bound states if their energy and the 
energy of scattering states used to approximate the $\Delta$ function satisfy the condition
  \begin{equation}\label{range}
|E|\le (Z_1Z_2e^2)^2\mu/2.
\end{equation}
As is noted in \cite{Sparen2}, the right-hand side of (\ref{range}) is just the nuclear Rydberg energy: 1 Ry$=(Z_1Z_2e^2)^2\mu/2$.
For  systems $d+\alpha$ and $\alpha+^{12}$C considered in \cite{BKMS2}  1 Ry = 0.13 MeV  and  10.7 MeV, respectively. These values clearly illustrate the conclusion made in \cite{BKMS2} that the $\Delta$ method  is quite appropriate for $\alpha+^{12}$C but fails for $d+\alpha$ due to a very narrow range of allowed energy values.  

\smallskip
\textbf{Inference.} In the strict mathematical sense the $\Delta$ method is not an analytic continuation of the denominator of the amplitude $\tilde f$ from the region $E>0$ to the region $E<0$, but it can still be used for practical purposes for sufficiently large charges and masses of colliding particles. The assertion about a strict mathematical proof of the correctness of the $\Delta$ method  \cite{OrIrNa} is incorrect.  This inference agrees with the results obtained in \cite{BKMS2,Sparen2}. 

The $\Delta$ method was used in \cite{IrOr1} to obtain ANCs for resonant nuclear states. In this regard, we would like to note that no special methods are needed for this purpose. Both the ERF  and $\Delta$ methods  were introduced to overcome the problem of the Coulomb singularity at $E=0$. However, the Coulomb-nuclear scattering amplitude does not possess the Coulomb singularity in the vicinity of resonances. Hence, one can simply continue analytically $\cot\delta_l$ from the real positive half-axis of $E$ to the resonance pole. 

\section{The $p-^{16}$O system}
Consider the  $p-^{16}$O system.  For this system  $m_1=m_p$=938.272 MeV, $m_2=m_{^{16}\mathrm{O}}$=14895.079 MeV, $Z_1Z_2$=8. 
$^{17}$F nucleus has two bound states: the ground state $5/2^+$ ($l$=2) and the excited state $^{17}F^*(0.4953\,\mathrm{MeV};1/2^+)$, $l$=0.
 The binding energies $\varepsilon$ of $^{17}$F(ground) and $^{17}$F*(0.4953 MeV) in the  $p-^{16}$O channel are 0.6005 MeV and
0.1052 MeV, respectively \cite{F17}.

In this section we present the proton  ANCs of ${}^{17}{\rm F}$ for the first excited state and for the ground state  obtained by extrapolation of the ERF and $\Delta$ functions to the bound state poles of ${}^{17}{\rm F}$.  They should be compared with the  experimental proton ANCs $C_{0}$   
 for the virtual decay   ${}^{17}\mathrm{F}\to ^{16}\mathrm{O}(2s_{1/2^{+}})  +p$  and $C_{2}$  for the virtual decay   
 ${}^{17}\mathrm{F}\to ^{16}\mathrm{O}(1d_{5/2^{+}})  +p$
 shown in  Table \ref{tableANC}. These ANCs are obtained from analyses of the astrophysical $S_{1\,16}$-factors \cite{Artemov}, the peripheral proton transfer reactions 
 populating the ground and excited states of ${}^{17}{\rm F}$ \cite{Gagliardi,Artemov96} and the radiative capture 
$^{16}{\rm O}(p,\gamma)^{17}$F reaction \cite{HBG10}.  
%The table also shows $C_{0}$ obtained from analysis of the elastic scattering data parametrized using the effective field theory (EFT) \cite{EFT}.
The table also shows $C_{0}$ determined  from fitting the effective field theory (EFT) S-factor to the experimental one \cite{EFT}. 
Below we explore the extrapolation of the elastic scattering data to the bound states of ${}^{17}{\rm F}$  to obtain the proton's ANCs   of its excited  and ground states. We demonstrate  that  the addressed here method of the extrapolation of the elastic scattering data  to the negative energy region
can be considered as another very useful practical method to extract the  ANCs from the experimental data. 

%excited state and $C_{2}$ for the ground state of $^{17}$F are given.
\begin{table}[htb]
\caption{ The experimental proton ANCs $C_{0}$  for the  excited state  and  $C_{2}$ for the ground state of $^{17}$F.}
\begin{center}
\begin{tabular}{|c|c|c|}
\hline
 %& \multicolumn{3}{|c|}
  $\,C_{0}$,  fm$^{-1/2}$    &  $\,C_{2}$,  fm$^{-1/2}$  & Reference \\  
%\hline 
% fm$^{-1/2}$ &   fm$^{-1/2}$ &     \\

 \hline
  $75.5 \pm 15$ & $1.1 \pm 0.33$ &  \cite{Artemov} \\
 $81 \pm 26$  & $1.1 \pm 0.10$ & \cite{Gagliardi} \\
 $73.0$ & $1.0 $ & \cite{Artemov96} \\
 $ 77.21$ & $0.91 $ & \cite{HBG10} \\
 $ 79.3 \pm 3.9$ &  & \cite{EFT} \\

\hline
\end{tabular}
\end{center}
\label{tableANC}
\end{table}  

The proton ANCs of ${}^{17}{\rm F}$  were also calculated using various theoretical approaches, see, for example, \cite{BDH98, nazarewicz}. In particular, the results of microscopic calculations \cite{BDH98} are as follows: $C_0$=91.14 
fm$^{-1/2}$, $C_2$=0.97 fm$^{-1/2}$ for the V2 potential and $C_0$=86.42 
fm$^{-1/2}$, $C_2$=1.10 fm$^{-1/2}$ for the MN potential.

 According to \cite{Sparen2,BKMS2}, the larger the charges and masses of colliding particles, the less the error associated with the use of the $\Delta$ method. The numerical parameter, which characterizes the accuracy of the $\Delta$ method, is the value of the Rydberg energy of the given system (see Section III above). For the $p-^{16}$O system 1 Ry$=1.50358$ MeV. This value is between the values 0.13 MeV and 10.7 MeV corresponding to the Rydberg energies for the $d+\alpha$ and $\alpha+^{12}$C systems, respectively. Remind that the $\Delta$ method turned out to be quite successful for $\alpha+^{12}$C but failed for $d+\alpha$ \cite{BKMS2}.
 
The ANC is obtained by analytic approximation of the ERF and $\Delta$ function by polynomials in $E$ and the subsequent analytic continuation  of these polynomials to the negative energy region. The coefficients of the polynomials are determined by the $\chi^2$ method using the experimental  phase shifts for $p-^{16}$O elastic scattering. To ensure the correct experimental position of a bound-state pole, 
the values of ERF and $\Delta$ function at $E =-\varepsilon$ are added as fitting parameters to their values at positive energies: 
$\,K_l(E)|_{E =-\varepsilon} = 2 \eta k^{2l+1} h(\eta)v_l(\eta)|_{E =-\varepsilon}$, $\;\;\Delta_l(E)|_{E =-\varepsilon} = 0$.

To employ the $\chi^2$ criterion, the errors equal to 
$\pm 1^{\circ}$ are applied to phase shifts $\delta_l(E)$. If $\delta_l+1^{\circ}$ exceeds  $180^{\circ}$, the value $179.99999999^{\circ}$ is used instead of $\delta_l+1^{\circ}$.  We use Eqs. (\ref{res2}) and (\ref{res22}) to find the ANCs.

\subsection{ANC for the excited state of $^{17}$F}

We begin with the analysis of the $1/2^+$ state of the $p-^{16}$O system ($l=0$).  For this state, we use the results of the latest   phase shift analysis obtained in Ref. \cite {Dubov}, in which 16 values of $\delta_0$  in the range of $E$= 0.3628 -- 1.8738 MeV   are presented.  
First, let us consider the approximation of the ERF  $\,K_0(E)$.  Our calculations are presented in the 2nd and 3rd columns of Table \ref{table1}. In this table, as well as in the following Table \ref{table2}, $\,N\,$ denotes the power of the approximating polynomial.  One sees that the obtained  ANC $\,C_0\,$  is convergent with increasing $N$. Convergence is achieved  already with $N=3$. Hence we can  consider the variant $N = 3$ as sufficient. 
 
\begin{table}[htb]
\caption{ANC $C_0$ for the excited state of $^{17}$F.}
\begin{center}
\begin{tabular}{|c|c|c|c|c|}
\hline
 & \multicolumn{2}{c|}{ERF method} & \multicolumn{2}{c|}{$\Delta$ method} \\  
\hline 
N & $C_0$, fm$^{-1/2}$ & $\chi^2$ & $C_0$, fm$^{-1/2}$ & $\chi^2$ \\ 
\hline 
1 & 121.65596 & 0.070 & 54.06743 & 0.7911 \\
2 & 101.86426 & 0.061 & 89.13841 & 0.0789 \\
3 & 101.86559 & 0.065 & 89.13140 & 0.0846 \\
4 & 101.86559 & 0.070 & 89.13140 & 0.0911 \\
5 & 101.86559 & 0.076 & 89.13140 & 0.0986 \\
\hline
\end{tabular}
\end{center}
\label{table1}
\end{table} 

\begin{figure}[htb]
\center
\includegraphics[width=0.9\columnwidth]{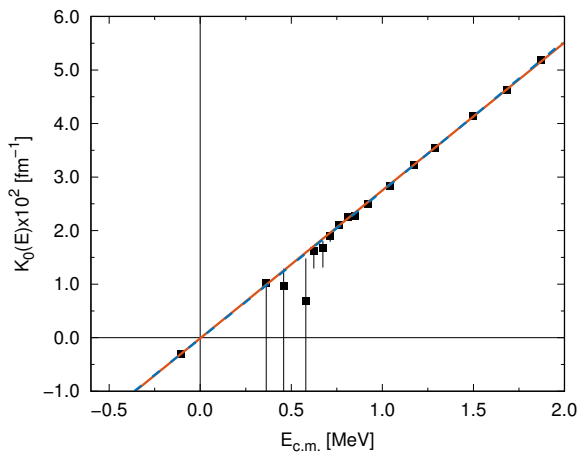}
\caption {Polynomial approximation of $K_{0}(E)$  for  $p+ ^{16}\mathrm{O}$ scattering  in the  $J^{\pi}=1/2^+$ state. Black squares with errorbars are the results obtained from the experimental scattering  phase shifts 
\cite{Dubov}. Solid red line is the polynomial approximation with $N=1$,  dashed blue line is the polynomial approximation with $N=2$. The lines corresponding to  higher $N$ are practically indistinguishable  from the $N=2$ line due to fast convergence. Therefore, they are not shown.}
\label{fig1}
\end{figure}

Figure \ref{fig1}  shows the polynomial approximation of the function $K_0(E)$. Note that the value of $K_0(E)$ at the origin is not zero but is very small and  can not be distinguished from zero in the scale of this figure.

\begin{figure}[htb]
\center
\includegraphics[width=0.9\columnwidth]{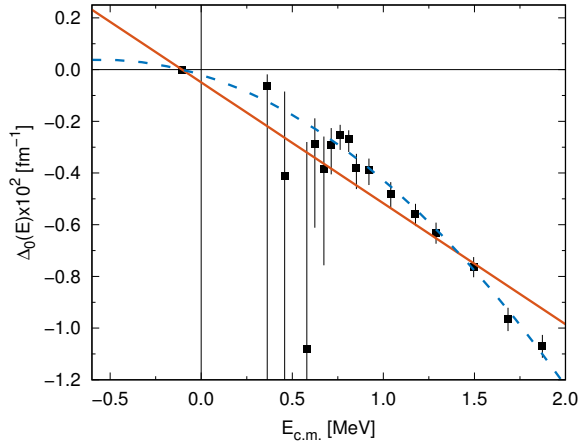}
\caption {Polynomial approximation of  $\Delta_0(E)$   for  $p+^{16}\mathrm{O}$ scattering  in the  $J^{\pi}=1/2^+$ state. The notations are the same as in Fig. \ref{fig1}.}
\label{fig2}
\end{figure}

Coinsider now the analytic continuation of the $\Delta$ function. Function $\Delta_0(E)$ is  approximated by polynomials in the same way as
for $K_0(E)$. The polynomial approximation of $\Delta_0(E)$ is shown in Fig.~\ref{fig2} and the results of the calculations are given in the 4th and 5th columns of Table \ref{table1}. 
It is seen that, similarly to the case of  $K_0(E)$, the  ANC $C_0$  converges rapidly with increasing $N$. Convergence is reached also with $N = 3$ and the result is  $\,C_0$= 89.13140 fm$^{-1/2}$.  This value  does not deviate much from $\,C_0$= 101.86559 fm$^{-1/2}\,$ obtained using polynomial approximation of  $\,K_0(E)$. The difference between these values can be related to the approximate nature of the $\,\Delta$ method. Note that the upper bound of the used energy interval ($E$=1.8738 MeV) slightly exceeds the value 1 Ry$=1.50358$ MeV for the $p-^{16}\mathrm{O}$ system. As was mentioned in Section III, 1 Ry can be considered as an upper bound for employing the $\Delta$ method. 
Note that the extrapolated ANCs are in a reasonable agreement with the experimental ANCs from Table \ref{tableANC}.

\subsection{ANC for the ground state of $^{17}$F}
\begin{figure}[htb]
\center
\includegraphics[width=0.9\columnwidth]{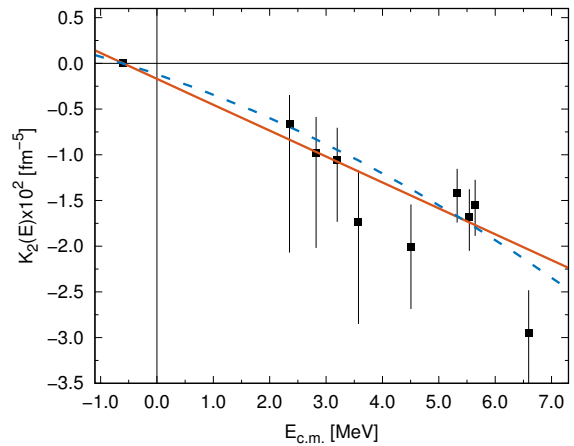}
\caption {Polynomial approximation of $K_2(E)$  for $p+ ^{16}\mathrm{O}$ scattering  in the  $J^{\pi}=5/2^+$ state. Black squares with errorbars are the results obtained from the experimental scattering  phase shifts 
\cite{Blue}.  Other notations are the same as in Fig. \ref{fig1}.}
\label{fig3}
\end{figure}

Owing to the absence of more recent phase shift analyses of $p+^{16}$O scattering in the $5/2^+$ state, we use here the rather old results of the phase shift analysis \cite{Blue} in which 9 values of $\delta_2(5/2)$  in the interval of $E= 2.35 - 6.60$ MeV  were presented. The procedure is analogous to the one used for the excited state described above. The corresponding ANC is denoted by $C_2$.  The results of the polynomial approximation of the ERF are shown in the 2nd and 3rd columns of Table \ref{table2}  and in Fig. \ref{fig3}.    

\begin{table}[htb]
\caption{ANC $C_2$ for the ground state of $^{17}$F.}
\begin{center}
\begin{tabular}{|c|c|c|c|c|}
\hline
 & \multicolumn{2}{c|}{ERF method} & \multicolumn{2}{c|}{$\Delta$ method} \\  
\hline 
N & $C_2$, fm$^{-1/2}$ & $\chi^2$ & $C_2$, fm$^{-1/2}$ & $\chi^2$ \\ 
\hline 
1 & 0.71537 & 0.16 & 0.52260 & 0.36 \\
2 & 0.87884 & 0.18 & 2.35850 & 0.19 \\
3 & 0.87881 & 0.20 & 2.33879 & 0.22 \\
4 & 0.87881 & 0.23 & 2.33876 & 0.26 \\
5 & 0.87881 & 0.28 & 2.33876 & 0.31 \\
\hline
\end{tabular}
\end{center}
\label{table2}
\end{table} 

It is seen that, similar to the case of the excited state of $^{17}$F,  the  ANC $C_2$  quickly converges with increasing $N$. The convergent result for ANC of $C_2=0.88$ fm$^{-1/2}$ is achieved with $N=3$.   Note that the ANC obtained using $K_{0}(E)$  polynomial approximation  is close to the 
%experimental 
ANCs from Table \ref{tableANC}. 

\begin{figure}[htb]
\center
\includegraphics[width=0.9\columnwidth]{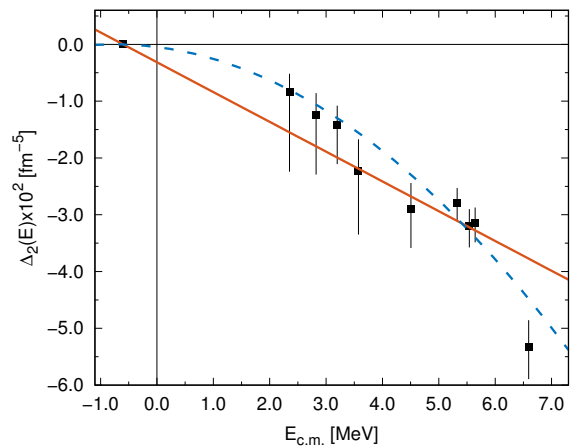}
\caption {Polynomial approximation of $\Delta_2(E)$ for $p+^{16}\mathrm{O}$ scattering in the state  $J^{\pi}=1/2^+$.  The notations are the same as in Fig. \ref{fig3}.}
\label{fig4}
\end{figure}

The results of the polynomial approximation of $\Delta_2(E)$ are shown in  3rd and 4th columns of Table \ref{table2}  and in Fig. \ref{fig4}.  Although the results appear to converge, however
they converge to an obviously wrong value. Most likely, this is  due to the energy  interval used for the approximation ($E$= 2.35 -- 6.60 MeV)  far exceeding the applicability limit of the $\Delta$ method for the $p-^{16}$O system of $1$ Ry$=1.50358$ MeV as discussed above.

\section{Conclusions}

 It is shown that  the $\Delta$ method suggested in \cite{Sparen}  is not strictly correct in the mathematical sense since it is not an analytic continuation of a partial-wave scattering amplitude  to the region of negative energies.  However, it can be used for practical purposes for sufficiently large charges and masses of colliding particles. It was demonstrated in the previous paper \cite{BKMS2} that this method was effective for the $\alpha-^{12}$C system ($Z_1Z_2=12$) but failed  for the $d-\alpha$ system ($Z_1Z_2=2$). In the present work, both the ERF and $\Delta$ methods of analytic continuation of scattering data are applied to the $p-^{16}$O system ($Z_1Z_2=8$) which can be considered as intermediate between $d-\alpha$ and $\alpha-^{12}$C systems. Both methods are used to determine the ANCs  for the ground 
$5/2^+$ and excited $1/2^+$ states of $^{17}$F nucleus in the $p-^{16}$O channel. 
Possible errors are added to experimental phase shifts. 

The values of the ANC $C_0$ for the excited $1/2^+$ state of $^{17}$F  obtained in the present paper on the basis of the phase-shift analysis of Ref. \cite{Dubov} are 101.9 fm$^{-1/2}$ and 89.1 fm$^{-1/2}$ for the ERF and $\Delta$ methods, respectively. They are not much different from each other. Note that both ANCs  are in a reasonable agreement with the 
experimental ANCs, see Table \ref{tableANC}.   The ANC $C_2$ for the ground state $5/2^+$ extracted using the phase-shift analysis of \cite{Blue} is 0.88 fm$^{-1/2}$ for the ERF method and 2.34 fm$^{-1/2}$ for the $\Delta$ method. The value 0.88 fm$^{-1/2}$ is close to the 
experimental  ANCs,  see Table \ref{tableANC}.  The polynomial approximation of $\Delta_{2}(E)$  for $p-^{16}\mathrm{O}$ scattering in the state  $J^{\pi}=5/2^+$  leads to the ANC $C_2=2.34$ fm$^{-1/2}$, which is  significantly higher than the range of this ANC available in the literature and should be considered as erroneous. Such a large discrepancy between the results of the ERF and $\Delta$ methods most likely is due to the fact that the energy interval used for the polynomial approximation of  $\Delta_2(E)$ function  far exceeds the limit of the applicability of the $\Delta$ method.
 
Summarizing, in this paper we demonstrated that the polynomial extrapolation of the  ERF and $\Delta$ functions  with the preset experimental binding energy gives converging and very reliable results for the proton ANCs of the ground and first excited states of  ${}^{17}{\rm F}$. We presented a  practical tool for experimentalists to determine the ANCs from the  measured  elastic scattering phase shifts. In nuclear astrophysics one needs to know the neutron ANCs. However, it is difficult to accurately measure the neutron elastic scattering phase shifts. Using the methods described here one can determine the proton  ANCs  from the proton elastic scattering phase shifts and then  using the mirror symmetry determine the neutron ANCs of the mirror nuclei \cite{TJM03,M12}. The same method can be used to determine the alpha-particle ANC on an unstable nucleus if the mirror alpha-particle ANC on a stable nucleus can be determined using elastic scattering data. 
Another very promising application of the extrapolation method addressed here is the effective field theory. In the EFT the elastic scattering data are analyzed  at positive energies and parametrized in terms of the EFT parameters \cite{Bogner,EFT}.  These parameters can be related to the EFR ones and can be used to extrapolate 
 the elastic scattering phase shifts to bound state poles to determine the ANCs \cite{EFT} .

\section*{Acknowledgements}

This work was supported by the Russian Science Foundation
Grant No. 16-12-10048 (L.D.B.) and the Russian Foundation
for Basic Research Grant No. 16-02-00049 (D.A.S.).
A.S.K. acknowledges a support from the Australian Research
Council. A.M.M. acknowledges support from the U.S. DOE
Grant No. DE-FG02-93ER40773, the U.S. NSF Grant No.
PHY-1415656, and the NNSA Grant No. DE-NA0003841.

\end{document}